\documentstyle[12pt]{article}
\begin{document}
\baselineskip=0.7cm
\newcommand{\EQ}{\begin{equation}}
\newcommand{\EN}{\end{equation}}
\newcommand{\EQA}{\begin{eqnarray}}
\newcommand{\EQN}{\end{eqnarray}}
\newcommand{\e}{{\rm e}}
\newcommand{\Sp}{{\rm Sp}}
\newcommand{\Tr}{{\rm Tr}}
\def\thefootnote{\fnsymbol{footnote}}
\makeatletter
\def\section{\@startsection{section}{2}{\z@}{-3.25ex plus -1ex minus 
 -.2ex}{1.5ex plus .2ex}{\normalsize\it}}
\makeatother

\begin{flushright}
hep-th/9703078\\
UT-Komaba/97-2
\end{flushright}

\vspace{1cm}

\begin{center}
\Large 
Schild  Action 
and 
Space-Time Uncertainty Principle 

in String Theory

\vspace{1.2cm}
\normalsize
{\sc Tamiaki Yoneya}
\footnote
{Internet address: {\tt tam@hep1.c.u-tokyo.ac.jp}
}

\vspace{0.5cm}
{\it Institute of Physics, University of Tokyo, Komaba, Tokyo}

\vspace{2.5cm}
Abstract\\
\end{center}
We show that the path-integral quantization of relativistic strings 
with the Schild action is essentially equivalent to the usual Polyakov  
quantization at critical space-time dimensions. 
We then present an interpretation of the Schild action
 which points towards a derivation  
of superstring theory as a theory of quantized space-time 
where the squared string scale, $\ell_s^2 \sim \alpha'$, plays the role of the 
minimum quantum for space-time areas. 
A tentative approach towards such a goal is proposed,  
 based on a microcanonical formulation of 
large $N$ supersymmetric matrix model. 

\newpage

\noindent
{\it Introduction}

Modern quantum theory of relativistic (super) strings usually starts   
from an action  which is 
quadratic with respect to the space-time 
coordinates $X^{\mu}$
\footnote{
The metric in the present note is 
Minkowskian $(+,+,\ldots, +, -)$ for both 
space-time and world sheet unless specified otherwise.},
\EQ
S_P = -{1\over 4\pi\alpha'}\int d^2\xi\, \sqrt{-g}g^{ab}{1\over 2}\partial_a X\cdot \partial_b X + \cdots.
\label{polyakovaction}
\EN
In contrast with older and more geometrically motivated 
Nambu-Goto action,
\EQ
S_{NG} =-{1\over 4\pi\alpha'}\int d^2\xi \,\sqrt{-\det (\partial _a X\cdot \partial_b X)} + \cdots,
\label{nambuaction}
\EN
the former is appropriate for the path-integral 
quantization of strings \cite{Polyakov}.  In fact, 
the former seems to be 
the unique action which is tractable in a direct path-integral 
approach.  Here and in what follows, by the ellipses we 
imply the additional terms which must enter to make the theory 
supersymmetric. 

There are an infinite number of 
different actions which are classically equivalent 
to the Nambu-Goto action. 
For example,  
\EQ
S_n =-\int d^2\xi\, e \Bigl\{{1\over e^n}
[-{1 \over 2\lambda^2}(\epsilon^{ab}\partial_a X^{\mu}\partial_b  X^{\nu})^2]^{n/2} + 
n-1\Bigr\} +\cdots
\EN
with arbitrary nonzero $n$ gives the same classical equations of 
motion, where 
the auxiliary field $e(\xi) \,  (>0) $ have the same transformation 
property as $\sqrt{-g}$ under the world-sheet reparametrizations, 
and $\lambda=4\pi\alpha'$.  The action $S_{NG}$ is a special limit 
with $n\rightarrow 1$,  where  
the auxiliary field decouples .  
Another special case which has been studied in the past is 
$n=2$, which is essentially the case first proposed by Schild \cite{Schild}.  
 
Some interesting remarks on the $n=2$ case 
have been made by Nambu \cite{Nambu}. In particular, he suggested 
an interesting interpretation for the quantity 
$
\sigma^{\mu\nu} \equiv 
\epsilon^{ab}
\partial_a X^{\mu}\partial_b X^{\nu}
$
as a formal analog of the Poisson bracket, and further speculated 
a quantization of the world-sheet by regarding 
 the space-time coordinates as a sort of gauge fields. 
His argument is a precursor of some of recent proposals  
concerning possible matrix representations
 \cite{Witten}\cite{BFSS} of the 
space-time coordinates, as has been motivated by string-theory dualities 
and D-branes \cite{Pol}. 
In particular, the Schild action plays an important role in a more 
recent proposal made in \cite{IKKT}. 
To the best of our knowledge, however,  
quantum string theory based on the Schild action 
has never been  concretely developed, 
except for an early attempt made in  
\cite{Eguchi}.
\footnote{
See a note added at the end.}

The purpose of the present note is to  show that 
the action $S_2$ is quantum-mechanically equivalent to the 
action $S_P$ at critical space-time dimensions, 
and then to discuss its relevance 
for a formulation of a space-time uncertainty relation 
which has been proposed \cite{yo} as a qualitative 
characterization of the  
space-time structure in fundamental string theory.  
We have recently argued \cite{liyo} that the space-time 
uncertainty relation of \cite{yo} and \cite{yo2}  nicely fits the 
short distance structure of 
the D-particle dynamics \cite{dparticle}. 
We will emphasize a 
view point that the 
above Poisson structure 
is a manifestation of conformal structure of the Polyakov 
quantization and is related to a possible realization 
of space-time uncertainty principle. 
Based on these observations, we will propose a tentative approach 
towards nonperturbative definition of string theory 
in terms of a microcanonical matrix model.  

\vspace{0.2cm}
\noindent
{\it Equivalence of the Schild and Polyakov quantizations}

Classical equivalence between the actions $S_n$ and $S_{NG}$ 
is trivially established by using 
the variational equation for the auxiliary field $e$
\EQ
 {1\over e}\sqrt{-\sigma^2}=\lambda,  \quad \sigma^2 \equiv {1\over 2}(\sigma^{\mu\nu})^2
\label{conformalconstraint}
\EN
which leads to $S_n=nS_{NG}$.  We will call this equation as  
the conformal constraint for a reason that will become clear below.  
In the Hamiltonian formalism, the conjugate momenta to 
the world sheet coordinates $X^{\mu}$ 
are 
\EQ
{\cal P}^{\mu} = -{e\over \sigma^2}({\dot X}^{\mu}({\acute X})^2 
-{\acute X}^{\mu}({\dot X}\cdot{\acute X})).
\label{conjugatemomenta}
\EN
Here of course the dot (${\dot X}$) and the prime (${\acute X}$) are 
derivatives with respect to  the world-sheet time and 
space coordinates, respectively. 
Combined with the conformal constraint (\ref{conformalconstraint}), 
it is easy to see that the momenta satisfy 
the usual classical Virasoro conditions
\EQA
{\cal P}^2 + {1\over \lambda^2}{\acute X}^2 &=& 0 ,\label{virasoro1}\\
{\cal P}\cdot {\acute X} &=&0 .\label{virasoro2}
\EQN
The Virasoro conditions in general express the conformal invariance 
of the world sheet theory. When we start from  the Schild-type action, 
they are thus 
a consequence of the equation (\ref{conformalconstraint}) (hence, 
the name ``conformal constraint"). 
More precisely, the conformal constraint is responsible only for the 
first Virasoro condition (\ref{virasoro1}) and the second one 
follows from the definition (\ref{conjugatemomenta}) alone. 
This reflects the fact that the second condition is 
kinematical in its nature: It only represents  invariance 
of the theory under an 
arbitrary reparametrization of the string at 
fixed world-sheet time. 
Conversely, under the kinematical condition (\ref{virasoro2}), 
the first constraint (\ref{virasoro1}) 
necessarily leads to the conformal constraint 
(\ref{conformalconstraint}) since the general 
solution to (\ref{virasoro2}) takes the 
form $-a(\xi)({\dot X}^{\mu}({\acute X})^2 
-{\acute X}^{\mu}({\dot X}\cdot{\acute X}))$ 
for some scalar function $a(\xi)$, 
leading to the identification $a=1/e$ on  
imposing the first condition (\ref{virasoro1}). 
Given these primary constraints (\ref{virasoro1}) and ({\ref{virasoro2}), 
following Dirac's procedure after introducing two 
lagrange-multiplier fields corresponding to the 
non-trace part of the world sheet metric in 
(\ref{polyakovaction}), we arrive at  
the same Hamiltonian as for the quadratic action $S_P$. 
This establishes equivalence of 
three different actions at the formal classical level. 

It is not obvious that classical equivalence in the above sense 
automatically leads to equivalence quantum mechanically. 
A typical argument for a quantum mechanical derivation 
of the quadratic action starting from the Nambu-Goto 
action is given in \cite{Polyakovbook} which generically 
shows that the former is obtained after making a rescaling for 
the string coordinates $X^{\mu}$ appropriately by 
a divergent constant and fine-tuning the cosmological 
constant. In general,  
 quantum string theories based on the Nambu-Goto 
action requires to make renormalization 
of the string-theory constants such as the tension and the 
cosmological constant ($\sim $ tachyon condensation). 
As is well known by now, a 
precise control over this problem is only available for  
systems with one or lower target space-time dimensions, 
which are most conveniently  
formulated in terms of old matrix models with $c\le 1$. 

In the case of the Schild action with $n=2$, we can 
derive the Polyakov action directly 
without making connection with the Nambu-Goto action. 
To see this, let us introduce  
auxiliary fields $t^{12}(=t^{21}), t^{11}, t^{22}$, which form  
a tensor density of weight two and rewrite the action   
\EQ
S(t,e,X)= \int d^2\xi \, {1\over e}[\det t 
-{1\over \lambda} t^{ab}\partial_aX \cdot \partial_bX ] -\int d^2\xi \, e.
\EN
This action is equivalent to the original Schild action $S_2$, 
even quantum mechanically, since the 
difference is a quadratic form of the auxiliary fields:
\EQ
S_2=S(t, e, X)  - \int d^2\xi  {1\over e}\det \tilde{t}
\label{gaussiancompletion}
\EN
where  
\EQ
\tilde{t}^{ab}= t^{ab}- {1\over \lambda}\epsilon^{ac}\epsilon^{bd}
\partial_c X\cdot \partial_d X.  
\EN
Let us then make a change of variables 
$
t^{ab}\rightarrow g^{ab}, 
e \rightarrow \tilde{e}
$
where $g_{ab}$ transforms 
as the standard world-sheet metric and $\tilde{e}$ as
a scalar,  
$ t^{ab}= g^{ab}e^2, e = \tilde{e}\sqrt{-g}$. 
Then the action reduces to 
\EQ
S(t,e,X)=  \int d^2 \xi (\tilde{e}^3-\tilde{e}) \sqrt{-g} -
{1\over \lambda}\int d^2\xi \tilde{e}\sqrt{-g}g^{ab}
\partial_aX\cdot \partial_bX
\label{zexg}
\EN

At this point, we can assume that the measure 
in the partition function has been defined 
 such that the total integration measure 
$[d\tilde{e}][dX][dg]$ obtained in these transformations is  
reparametrization invariant. 
We then decompose the measure $[dg]$ as usual 
$$
{[dg][dX]\over [d({\rm diff}_2)]}=
[d{\rm Weyl}]{[dg][dX]\over [d{\rm Weyl}] [d({\rm diff}_2)]}
$$
to separate the Weyl mode.  When the conformal anomaly 
which generically appears in this process 
 cancels, we see that the conformal mode 
is contained only in the first term of   (\ref{zexg}). 
Thus at critical space-time dimensions, we have the equation of 
motion 
$
\tilde{e}^3-\tilde{e}=0
$
from the variation of the conformal factor. 
Only allowed solution of this equation is $\tilde{e}=1$ 
since we assume that $e>0$.  Thus 
the action (\ref{zexg}) reduces to the 
quadratic aciton, and the quantization of the Schild action $S_2$ is 
essentially equivalent to the Polyakov quantization. 

Rigorously speaking, however, we have to be more 
careful with respect to the definitions of the continuum  
path-integrals, including more precise 
(regularized) definition of the measure and, in particular,  the ranges of various auxiliary 
fields.  Despite these subtleties, 
we propose to take the above observation as a hint for  
a new interpretation of string theory towards its nonperturbative 
forumulation. 

\vspace{0.2cm}
\noindent 
{\it Conformal constraint and  space-time uncertainty principle}

We now want to clarify the meaning of the conformal constraint 
(\ref{conformalconstraint}). 
It was already emphasized that, 
from the view point of the Schild action,
 conformal invariance of world-sheet 
string theory is a consequence of the 
conformal constraint.  
Now, what is the most crucial physical property 
of string theory resulting from the world-sheet conformal invariance? 
In some of earlier works \cite{yo}\cite{yo2}, we have 
proposed a possible answer for this question that the conformal 
invariance is related with a space-time uncertainty principle of the 
form 
\EQ
\Delta X \Delta T \sim \lambda
\label{stuncertaintyrelation}
\EN
for the minimum uncertainties with respect to the measurements of the 
lengths of the  space ($\Delta X$) and time ($\Delta T$) 
intervals
\footnote{
There has been other proposals for extended uncertainty relations 
\cite{amelio} motivated 
from string theory. }.  
If we remember the reparametrization invariant 
Poisson bracket structure
\EQ
\{X^{\mu}, X^{\nu}\} \equiv {1\over e}
\epsilon^{ab}\partial_a X^{\mu}\partial_b X^{\mu}, 
\label{poissonbracket}
\EN
the conformal constraint takes the following very suggestive form 
\EQ
-{1\over 2}(\{X^{\mu}, X^{\nu}\})^2 = \lambda^2. 
\label{conformalconstraint2}
\EN

To exhibit the relation between (\ref{stuncertaintyrelation}) 
and (\ref{conformalconstraint2}), let us have a  recourse to 
a simple qualitative example considered in \cite{yo2}. 
We consider an amplitude for the 
mapping from a rectangular region of the world sheet 
with the side lengths $a, b$ in the conformal gauge to 
a corresponding rectangular region of the target  space-time 
whose lengths are $A$ and $B$ in the time  ($\mu=0$) and 
the space ($\mu=1$) directions, respectively. 
The boundary conditions for the space-time 
coordinates are 
\EQ
X^{\mu}(\xi^0, 0)=X^{\mu}(\xi^0, b)=\delta^{\mu 0}{A\xi^0\over a},
\EN
\EQ
X^{\mu}(0, \xi^1)=X^{\mu}(a, \xi^1)=\delta^{\mu 1}{B\xi^1\over b}.
\EN

Then the quadratic action in the Euclidean metric 
gives the following factor for the 
amplitude, apart from a power behaved measure factor 
which is irrelevant for the present qualitative discussion, 
\EQ
\exp [-{1\over \lambda} (A^2{b \over a} +B^2{a\over b})]. 
\label{wavepacket}
\EN
The ratio $a/b$ is nothing but the unique conformal invariant 
(namely, modular 
parameter) for the rectangle on the world sheet 
which is known as the ``extremal length" \cite{alfors}. 
 We note that two forms, $\Gamma=a/b$ or $\Gamma^*=b/a$, of the extremal 
lengths are nothing but the duality relation between the extremal 
length $\Gamma$ and its conjugate extremal length $\Gamma^*$ satisfying $\Gamma\Gamma^*=1$. 
The expression (\ref{wavepacket}) 
clearly shows that there is an uncertainty relation 
of the form (\ref{stuncertaintyrelation}) 
with 
$
\triangle T \sim \langle A \rangle 
\sim \sqrt{\lambda\Gamma} , \, \, \triangle X \sim \langle B \rangle
\sim \sqrt{\lambda\Gamma^*}. 
$.  

Now, making the  correspondence 
$(\lambda, A, B, \sqrt{a/b}, \sqrt{b/a}) 
\leftrightarrow (\hbar, x, y, \Delta x/\sqrt{\hbar}, \Delta p/\sqrt{\hbar})$, 
the form 
 (\ref{wavepacket})  is very analogous to   
Wigner's 
phase space representation of a  
density operator  ${\cal O}_{\Delta x}=|g(\Delta x)><g(\Delta x)|$ 
$$
{\cal O}(x,p)\equiv 
\int \, dy \, \e^{ipy \over \hbar}\, 
<x-{1\over 2}y|\, {\cal O}_{\triangle X}\, |x+{1\over 2}y>\, 
\propto \exp -\Big[({x\over \Delta x})^2 + ({p\over \Delta p})^2\Bigr],
$$
representing a Gaussian wave packet state $|g(\Delta x)>$
in ordinary particle quantum mechanics, where 
$<x|g(\Delta x)>\sim \exp -{1\over 2}({x\over \Delta x})^2$ and 
$\Delta x \Delta p = \hbar$ . If we further suppose to 
integrate out the density operator 
with respect to  the parameter $\Delta x$ in analogy with the 
string modular integral, and take the classical limit 
$\hbar \rightarrow 0$,  the integral is dominated by the 
classical solution $|{x\over p}| = |{\Delta x\over \Delta p}|$.  
This relation can be regarded as the classical counter part 
of the Heisenberg uncertainty relation, 
corresponding to the 
Poisson bracket relation $\{x, p\}=1$. The integration over the 
parameter $\Delta x$ of course means that we are now considering  
a statistical density operator. 

On the other hand, if we calculate the same amplitude 
using the Schild action in the gauge $e=1$, we have 
the area law factor
\EQ
\exp -({1\over \lambda^2}{(AB)^2\over a'b'}+a'b')
 \Rightarrow \exp -{2\over \lambda}AB
\label{transformationfunction}
\EN 
under the conformal constraint
\EQ
{AB\over a'b'}=\lambda,  
\label{abconstraint}
\EN
or more precisely, after integrating over the 
parameter $s\equiv a'b'$ 
making the change of variable $s \rightarrow \sqrt{s} -
{AB\over \lambda\sqrt{s}}$.  Here we put prime on the 
world-sheet length parameters to discriminate them from 
those of the quadratic action. 
The argument of the previous section requires that we must have 
the same result as for the quadratic 
action after integrating over the modular parameter ${a\over b}$. 
This becomes obvious by  making a correspondence of the 
integration parameters
$
a \leftrightarrow a' , \, \, b \leftrightarrow B^2/ (\lambda b')
$
by which the conformal constraint (\ref{abconstraint}) is reduced to 
${A\over B}={a\over b}$, the ``classical 
solution" for (\ref{wavepacket}). 
Thus, the conformal constraint can be 
interpreted as the classical 
counter part for the space-time uncertainty relation (\ref{stuncertaintyrelation}),  
in the same sense as in  the ordinary classical limit 
$\hbar \rightarrow 0$ for the  statistical density operator 
given by
$
\rho \equiv \int d(\Delta x) \, {\cal O}_{\Delta x}.
$
 
These elementary considerations strongly suggest the existence of 
a theory in which 
the Poisson structure (\ref{poissonbracket}) is  
replaced by a commutator between coordinate operators 
$
\{ X^{\mu}, X^{\nu}\} \rightarrow [X^{\mu}, X^{\nu}] ,
$ 
and also that such a theory should take into account 
some form of quantum condition which 
reduces to the conformal  
constraint  (\ref{conformalconstraint2}) 
in a certain classical  limit.  
Our discussion above shows that, in this 
kind of theories of quantized space-time, 
the ordinary space-time 
continuum should be 
 interpreted as something analogous to 
the classical phase space in quantum particle mechanics. 
A strong form of possible quantum conditions
\footnote{
A similar quantum condition has 
been considered in ref. \cite{doplicher} 
in an even more stronger form in 4 dimensions 
and rigorous inequalities similar to our uncertainty 
relations have been proved. 
I would like to thank 
S. Doplicher and A. Jevicki for bringing this 
work into my attention.  
} would be 
to demand that the operator $[X^{\mu}, X^{\nu}]$ 
satisfies 
\EQ
-{1\over 2}([X^{\mu}, X^{\nu}])^2=\lambda^2 I
\label{quantumconstraint}
\EN
where $I$ is the identity operator in some operator algebra 
and we assume the Minkowski metric, 
$$
{1\over 2} ([X^{\mu}, X^{\nu}])^2 \equiv 
-\sum_{i=1}^9 [X_0, X_i]^2 + {1\over 2} \sum_{i, j=1}^9 
[X_i, X_j]^2 .
$$
A weaker form of the 
quantum condition would be to 
demand (\ref{quantumconstraint}) only for 
appropriately defined expectation values 
$\langle \cdots \rangle$. 
This leads to the following inequality, 
\EQ
\langle ([X^0, X^i])^2\rangle  \, \, \ge \lambda^2, 
\EN
 and hence is consistent with  the space-time uncertainty 
relation (\ref{stuncertaintyrelation}) in a way that is invariant 
under Lorentz transformations.  Note that, without further 
conditions, no definite 
lower bound exits for the 
space-space components. This conforms to  the analysis in \cite{liyo} 
in string theory 
based on the results \cite{dparticle} on the dynamics of D-particles.  

\vspace{0.2cm}
\noindent
{\it Microcanonical matrix model}
 
We have seen that a key feature of the 
conformal invariance is encoded in the space-time uncertainty principle.  
It is natural to postulate that non-perturbative string theory 
should be formulated  on the basis  
of the space-time uncertainty principle. 
We now want to briefly discuss a tentative  
approach towards such a goal. 
To motivate our proposal, 
we first remark a striking analogy of 
the above structure to the matrix algebra \cite{Witten} 
appearing as the zero dimensional reduction  
of 10-dimensional super Yang-Mills 
theory as  describing the effective weak-coupling 
dynamics of D-instantons 
(namely, Dirichlet-(-1) branes) in the type IIB superstring theory. 
Assuming the validity of the conjectured self-duality of 10-dimensional 
type IIB theory, it is natural to suppose that the 
microscopic space-time structure seen through the 
D-branes is basically the same as that seen in the 
elementary-string excitation picture. 

Our postulate then is to identify the algebra for the 
D-instanton coordinate matrices $X_{\mu}  \, \, ({\mu}=1, 2, \ldots, 10 )$ 
which are $N\times N$ hermitian (anti-hermitian for $\mu=10$) matrices 
describing $N$ D-instantons 
with the operator algebra discussed in the previous section. 
As is well known, these matrices are nothing but 
the Yang-Mills fields coupled to open strings at their 
end points where the D-instantons are located. 

The quantum condition is assumed to be 
\EQ
\langle {1\over 2}([X_{\mu}, X_{\nu}])^2 \rangle \equiv 
\lim_{N\rightarrow \infty}
{1\over N}\Tr {1\over 2}([X_{\mu}, X_{\nu}])^2=-\lambda^2 .
\label{matrixqcondition}
\EN
Here, $\langle \cdot\rangle$ denotes the expectation value 
with respect to the $U(N)$ trace as indicated. 
We adopt a weaker form of the possible quantum conditions 
and assume the large $N$ limit to include the case of arbitrary 
number of D-instantons. From the work \cite{dwhni}, 
we know that the parameter $1/N$ plays the role of 
Planck constant in relating the Poisson structure to the 
commutator. 
Then it is natural to define  the fundamental partition function as 
\EQ
Z=
\int \Bigl(\prod_{\mu=1}^{10} \, d^{\infty^2}X_{\mu}
\Bigr){\cal J}[X]
\delta(\langle {1\over 2}([X_{\mu}, X_{\nu}])^2 \rangle+\lambda^2)
\EN
where $d^{\infty^2}X_{\mu}$ is 
the large N limit of the 
standard $U(N)$ invariant Haar measure and  ${\cal J}[X]$ is an 
additional measure factor to be determined below. 

Let us consider how to determine the measure factor 
 ${\cal J}[X]$.  The quantum condition has the gauge symmetry 
under 
\EQ
X_{\mu} \rightarrow UX_{\mu}U^{-1} + a_{\mu}I
\EN
where $U\in$ U(N) and $a_{\mu}$ are arbitrary real constants. 
The measure factor should respect this gauge symmetry. 
A further requirement to make the theory sensible 
is that the partition function 
should satisfy the cluster property. 
That is, distant D-branes must behave 
independently except for 
the possible power-behaved long-range forces between them. 
A more stronger condition is that 
the partition function be consistent with the known behavior of the classical BPS 
saturated solutions. From the view-point of  
Yang-Mills theory reduced to a point, the classical solutions are  
characterized as the solutions of the bosonic 
equations 
\EQ
[X_{\mu}, [X_{\mu}, X_{\nu}]=0
\EN 
which are nothing but the 
variational equation for the quantum condition (\ref{matrixqcondition}). 
These solutions can be regarded as D-branes as 
composites of the D-instantons. 
Explicit solution for the case of D-strings 
(i.e., D-1 brane) are constructed 
in ref. \cite{IKKT}. The higher-dimensional cases have 
also been discussed in several recent works.
\footnote{See, e.g. \cite{higherdbrane}. 
In particular, \cite{li} has discussed the solutions emphasizing 
 close affinity between the properties of the solutions and the space-time 
uncertainty relation.} 

The cluster 
property for distant D-branes requires that  the 
 fluctuations around 
the solutions cancel to the leading order. 
Otherwise the distant D-branes would in general interact through 
logarithmic long-range forces which cannot be 
accepted.   Together with the gauge symmetry, only conceivable 
way  to satisfy the 
above criteria  is to choose the measure factor ${\cal J}$   
such that the partition function be supersymmetric. 
The largest space-time dimensions in which this 
is possible is 10. This follows from a  
well known fact about supersymmetric 
Yang-Mills theory, after making the 
reduction to a point (i.e., reduction to $-1$ dimensions).  
Thus we are naturally led to the measure factor,  
\EQ
{\cal J}[X] = \int d^{16}\psi' \exp ({1\over 2}\langle \overline{\psi}
\Gamma_{\mu} [X_{\mu}, \psi]\rangle)
\EN
where $\psi$ is the $U(\infty)$-matrix whose elements are Majorana-Weyl 
spinors in 10 dimensions and the prime in the integration volume 
denotes  that possible fermion zero-modes should be removed for the 
partition function. 
The supersymmetry is easily established after  rewriting  
the partition function by introducing an auxiliary constant multiplier $c$, 
\EQ
Z=\int \, dc \, \Big(\prod_{\mu}d^{\infty^2}X_{\mu}\Bigr)
d^{16}\psi'  
\exp\Bigl[
c(\langle {1\over 2}([X_{\mu}, X_{\nu}])^2 \rangle+\lambda^2) + 
{1\over 2}\langle \overline{\psi}
\Gamma_{\mu} [X_{\mu}, \psi]\rangle\Bigr] .
\label{microcanonicalmtheory}
\EN 
The action in this expression is  invariant under 
two supersymmetry transformations, 
\EQA
\delta_{\epsilon}\psi&=&ic [X_{\mu}, X_{\nu}] \Gamma_{\mu\nu}\epsilon , 
\\
\delta_{\epsilon}X_{\mu}&=&i\overline{\epsilon}\Gamma_{\mu}\psi ,
\\
\delta_{\epsilon}c&=&0 ,\\
\delta_{\eta} \psi &=& \eta ,\\
\delta_{\eta} X_{\mu}&=&0 ,\\
\delta_{\eta}c&=&0 .
\EQN 
 
This looks  similar to the model considered in   
ref. \cite{IKKT} which we call the IKKT model 
except for an additional auxiliary 
variable $c$ which enters to realize  
 the  space-time uncertainty principle in a weak form. 
We now explain some features of our model. 
 First we shall argue that the IKKT model can 
be interpreted  from our model as an effective theory for D-branes.

Let us first consider the effective action 
for many distant clusters of D-brane systems. 
This corresponds to introducing the background $X_{\mu}^b$ in the 
block-diagonal form 
\EQ
X_{\mu}^b \equiv \pmatrix 
{Y_{\mu}^{(1)} & 0 &0 &  . & .  & .  \cr 
0 & Y_{\mu}^{(2)} &0 & .  & . & . \cr 
0 & 0 & Y_{\mu}^{(3)} & . & . & . \cr 
. & . & . & . & . & . \cr
. & . & . & . & . & . \cr
. & . & . & . & . & . \cr}
\EN
where each block is assumed to be $N_a\times N_a$ hermitian 
matrix. The separation between the clusters $Y_{\mu}^{(a)}$ 
and $Y_{\mu}^{(b)}$
 is measured by 
$\ell_{a, b}\equiv |{1\over N_a}\Tr_a Y_{\mu}^{(a)} 
-{1\over N_b}\Tr_b Y_{\mu}^{(b)}|$ which are here assumed to be 
sufficiently large compared to $\sqrt{\lambda}$.  The notation 
$\Tr_a$ means the trace operation within each block. 
The backgrounds must satisfy the conditions 
\EQ
-\sum_{a=1}^n \Tr_a {1\over 2}[Y_{\mu}^{(a)}, Y_{\nu}^{(a)}]^2 =N\lambda^2,
\label{condition1}
\EN
\EQ
\sum_{a=1}^n \, N_a =N \, \, (\rightarrow \infty).
\label{condition2}
\EN

Now let us suppose to evaluate the fluctuations around 
the backgrounds by 
decomposing as $X_{\mu}=X_{\mu}^b+ \tilde{X}_{\mu}$. 
For fixed $c$, the calculation is entirely the same as the 
IKKT model \cite{IKKT} where it is shown that the 
leading order one-loop effective action for the backgrounds 
decreases as $O({1\over \ell_{a,b}^8})$ in the limit of 
large separation. To one-loop order, the effective 
action is independent of the parameter $c$. 
Thus  the distant D-brane systems can be treated as independent 
objects in this approximation, and hence we can take into account the 
conditions (\ref{condition1}), (\ref{condition2}) in a 
statistical way by introducing two Lagrange 
multipliers $\alpha$ and $\beta$. Namely, we 
can derive the effective action for the 
D-brane sub systems by the same argument as 
we use in going from the microcanonical ensemble to 
the canonical distribution for subsystems. 
Then the effective partition function $Z_{\rm{eff}}$ for the D-brane 
sub system described $Y_{\mu}$ within the 
semi-classical approximation is given by a grand canonical form 
\EQ
Z_{\rm{eff}}= \sum_N \Bigl(\prod_{\mu}\int \, d^{N^2}Y_{\mu}
\Bigr) d^{16}\psi
\exp S[Y, \psi, \alpha, \beta],
\label{grandcanonicalform}
\EN
\EQ
S[Y, \psi, \alpha, \beta] \equiv  -\alpha N -\beta \Tr_N{1\over 2}[Y_{\mu}, Y_{\nu}]^2
-{1\over 2}\Tr \, \overline{\psi}[\Gamma_{\mu}, Y_{\mu}]\psi ,
\EN
where by $N$ we denote the order of the background submatrix $Y$ 
and $\Tr_N$ is the corresponding trace.  This 
form is identical with the IKKT model, except that in our case 
the following condition must be satisfied,  
\EQ
-\langle\langle  {1\over 2}[Y_{\mu}, Y_{\nu}]^2 \rangle\rangle 
\equiv -{\sum_N \Bigl(\prod_{\mu}\int \, d^{N^2}Y_{\mu}\Bigr) d^{16}\psi
\, {1\over N}
\Tr_N {1\over 2}[Y_{\mu}, Y_{\nu}]^2 \exp S[Y, \psi, \alpha, \beta]
\over \sum_N \Bigl(\prod_{\mu}\int \, d^{N^2}Y_{\mu}\Bigr) d^{16}\psi 
\exp S[Y, \psi, \alpha, \beta]} \nonumber \\
 =  \lambda^2
\EN
in order to respect the quantum condition. 
Note that the constant $\beta$ can be identified with the 
mean value of the original Lagrange multiplier $c$ in 
(\ref{microcanonicalmtheory}). 
The Lagrange multipliers can in principle be determined  
by requiring the equivalence with the microcanonical form  
in the present approximation.  The two constants of the effective theory should 
be related to the vacuum expectation values of 
scalar background fields of type IIB superstring theory, 
such as dilaton and/or  its dual partner, scalar axion.   
 
Our model is thus approximately equivalent with the 
IKKT model as an effective 
theory describing sufficiently distant 
D-brane systems in the 
nearly classical limit. 
However, we should not expect 
complete  equivalence in full quantum theory. 
First we note that the completely diagonal 
background is not a good background in the present case, 
except in the limit 
$\lambda \rightarrow 0$, since it cannot satisfy the 
quantum condition (\ref{condition1}). 
The quantum condition gives a rationale for 
choosing the solution satisfying the 
condition $[Y_{\mu}, Y_{\nu}] \propto {\cal I}$ 
for some pairs of the components $\mu, \nu$ where 
at least one pair must contain the time component and 
${\cal I}$ is a constant matrix whose square has a nonzero trace 
in the large $N$ limit.  
The authors of ref. \cite{IKKT} interpreted their model 
as an effective theory for the 
 reduced Yang-Mills theory in 10 dimensions. 
In perturbation theory, this interpretation 
suffers from nonrenormalizable infrared divergencies, 
reflecting nonrenormlizable ultraviolet difficulty of 
10 dimensional super Yang-Mills theory. 
Here  we would like to recall a remark \cite{stro} 
that the equivalence between the microcanonical and canonical 
formulations of non-renormlizable field theories 
is not at all obvious.  In a similar vein, our 
microcanonical matrix model 
is most probably not exactly equivalent with the reduced 
Yang-Mills theory. 
An obvious flaw of the grand canonical  
form  (\ref{grandcanonicalform}) as the  
 definition of the theory  
is that the action appeared in it is not bounded 
in the Minkowski metric. To make sense out of this form, 
it is necessary to make a Wick rotation 
$Y_0 \rightarrow iY_{10}$ and assume a 
positive value for $\beta$. 
It is not clear how this is justified within the logic of the 
present formalism. Note that the quantum condition in the 
form (\ref{matrixqcondition}) is 
meaningful only in the Minkowski metric.

\vspace{0.2cm}
\noindent
{\it Further discussions}

It seems that, through exploring the meaning of the 
Schild action in string theory, 
we have opened up innumerable new 
problems.   

First of all, the microcanonical form is not the unique possibility. 
We may try  to impose a stronger form of the quantum condition  
such as (\ref{quantumconstraint}) 
by generalizing the classical 
action $S_2$ with the auxiliary field $e(\xi)$.  
In that case, however, 
a naive introduction of an auxialiry  matrix 
corresponding to the variable $e$ in general 
violates the supersymmetry, except for the 
naive continuum theory in the large $N$ limit. 
Although such a model may still satisfy the 
cluster property in the one-loop approximation, 
lack of exact supersymmetry 
would make the theory 
ambiguous.  For example, 
if we cannot assume exact supersymmetry, 
it is not clear how to fix the measure ${\cal J}(X)$. 
To implement a stronger quantum condition, 
a higher symmetry than  [$U(N)$-gauge $+$ SUSY]  
 seems necessary, unless  we rely upon an argument assuming  
 universality even if we start from the 
non supersymmetric models in the continuum limit. 
Considering various other possibilities, our model  
should yet be regarded as a tentative working hypothesis 
as a first step for further exploration of 
space-time uncertainty relation as 
a fundamental principle underlying the string theory. 
We postpone more detailed investigations of the 
present model and other  
possiblities to forthcoming works. 
 
Secondly, one of the crucial questions is 
whether and in what sense the fundamental perturbative string theories 
are contained in the present formalism. 
If our interpretation that the matrix $X_{\mu}$ represents 
the D-branes is correct, it is natural to introduce 
the quantity $A(s)ds +\cdots \equiv X_{\mu} 
{\partial x^{\mu}(s) \over \partial s}ds +\cdots$ 
as the ``string-bit" operator, whose matrix element 
$A_{ab}(s)ds$ 
is an infinitesimal open segment of a bosonic 
string connecting two D-instantons labeled by $a$ and $b$,  
respectively, 
of which the infinitesimal
 space-time element is given by 
${\acute x}^{\mu}(s)ds$. 
In the perturbative string theory, this is nothing but the 
vertex operator for an open string connecting two D-instantons,  
and the dynamics of string coordinate $x^{\mu}(s)$ 
would encode the dynamics of the matrix variable $X_{\mu}$ 
as collective coordinates. The matrix interpretation 
of the D-brane coordinates has first been obtained 
in this picture \cite{Witten} as an effective 
low-energy theory. It is important to develop systematic 
formalism for deriving a full-fledged dynamical 
theory for the matrices along this line.  
Conversely, if the full dynamics of 
D-brane matrix $X_{\mu}$ were known, the dynamics of the 
fundamental strings should be derived from the former 
by treating in turn the string variables $x^{\mu}(s)$ 
as the collective coordinate. Then the 
operator representing a closed string configuration 
$x^{\mu}(s) \, \, (\, x^{\mu}(0)=x^{\mu}(2\pi)\, )$ would be 
the Wilson loop, ${1\over N}\Tr P\exp i (
{1\over \lambda}\oint ds X_{\mu}{\partial x^{\mu}(s)
\over \partial s})$ 
apart from some appropriate fermionic contribution, with $P$ being the 
ordering operator along the string
\footnote{
A slightly different suggestion has been made in \cite{IKKT} 
where ${\partial x^{\mu}(s)
\over \partial s}$ is replaced by the momentum density of the 
string. 
}.  
The derivation of string dynamics 
is now reduced to the long-standing 
question of deriving the 
string dynamics as effective theory from matrix models. 
This is essentially a dual transformation. 
Note that the commutator $[X_{\mu}, X_{\nu}]$ and 
the area derivative ${\delta \over \delta \sigma^{\mu,\nu}}$ 
corresponding to an infinitesimal area deformation 
$\delta\sigma^{\mu\nu}=\delta x^{\mu}\wedge\delta x^{\nu}$ 
of the Wilson loop 
are dual to each other in this dual transformation. 
At present, it is not clear whether the microcanonical 
formulation provides new handles on this old and difficult 
question. One thing evident  from the 
structure of the Schwinger-Dyson equation for the 
matrix model, however, is that the 
mean value of the Lagrange multiplier $c\sim \beta$ in 
the fundamental partition function is inversely proportional to the 
string coupling constant, $c \propto {1\over g_s\lambda^2}$,  
of string perturbation theory, 
being consistent with well known effective low-energy theories 
of D-branes \cite{Witten}. 
It should also be mentioned that under the dual 
transformation, the stronger quantum condition 
(\ref{quantumconstraint}) reduces  
to the Virasoro-type condition $(\delta/\delta \sigma^{\mu\nu})^2
\propto \lambda^2$,  whose classical 
counter part is (\ref{virasoro1}) as it should be,  
at least for a smooth 
string configuration $x^{\mu}(\sigma)$.  
This can be a further motivation in favor of the stronger 
quantum condition.  

Another important question is what is, if any, the relation 
of the present approach to the matrix model  (so-called , M(atrix) theory)
proposed in \cite{BFSS} as a nonperturbative light-cone 
formulation of the M-theory ($\sim$ type IIA superstring) 
which requires 11 (=10+1) dimensions. 
Our present approach requires 9+1 dimensions 
as a largest possible dimensions and 
conforms to the type IIB superstring.  
To connect to type IIA and/or 
M theory, 
we may compactify one dimension and 
invoke the T-duality, or may add one additional 
dimension. For the latter possibility, we might go over to 
the stochastic quantization of the model.  

Finally, a remaining crucial question is the problem of 
background independence. 
Ever since the first discovery that the string theory 
is a quantum theory of gravity \cite{schsch}\cite{yo0}, the geometrical 
formulation of string theory as a natural quantum extension 
of Einstein's general relativity has been a big 
mystery. In the present paper, 
we always assumed flat space-time background. 
It is not at all clear how to interpret the 
quantum conditions in a manner consistent with 
the principles of general relativity, which should ultimately be 
reconciled with the principles of string theory 
at least in the long distance regime. 
In this respect, it may be interesting to extend  
our interpretation of the conformal constraint 
to the case  with  non-trivial space-time background fields. 
The real issue, however, is how to understand the dynamics 
of space-time geometry itself (not just a different 
fixed curved space-time) within the present framework. 

Obviously much work remains to be done in order to 
realize the present proposals. 

\vspace{0.2cm}

I would like to thank A. Jevicki for sharing together 
an idea that D-instantons may be the fundamental 
building block for string theory and for collaborative 
discussions on this and related topics during my stay 
at Brown University in the summer in 1996, which was made 
possible under 
the US-Japan Collaborative Program for Scientific Research 
supported by the Japan Society for the Promotion of Science.  
Also I wish to thank my colleagues at Komaba, 
especially, M. Kato and Y. Kazama, 
for stimulating discussions on D-brane dynamics at our seminars. 

\vspace{0.2cm}
\noindent
{\it Note added}: 
After the paper first appeared on hep-th, Dr. Spallucci called 
my attention to  
the work \cite{ansoldi} which used the Schild action in a  
loop space approach to string quantization. 
Also, Dr. Tseytlin 
pointed out his work \cite{fradtseytlin} which established 
the equivalence of the Schild and Polyakov quantizations in the 
semi-classical approximation.

\end{document}